\documentclass[8.5pt,twoside,twocolumn]{article}
\usepackage[super,sort&compress,comma]{natbib} 
\usepackage{mhchem}
\usepackage{times,mathptmx}
\usepackage{sectsty}
\usepackage{balance} 

\usepackage[usenames]{color}

\usepackage{graphicx} 
\usepackage{lastpage}
\usepackage[format=plain,justification=raggedright,singlelinecheck=false,font=small,labelfont=bf,labelsep=space]{caption} 
\usepackage{fancyhdr}
\begin{document}

\thispagestyle{plain}
\renewcommand{\thefootnote}{\fnsymbol{footnote}}
\renewcommand\footnoterule{\vspace*{1pt}%
\hrule width 3.4in height 0.4pt \vspace*{5pt}} 
\setcounter{secnumdepth}{5}

\makeatletter 
\def\subsubsection{\@startsection{subsubsection}{3}{10pt}{-1.25ex plus -1ex minus -.1ex}{0ex plus 0ex}{\normalsize\bf}} 
\def\paragraph{\@startsection{paragraph}{4}{10pt}{-1.25ex plus -1ex minus -.1ex}{0ex plus 0ex}{\normalsize\textit}} 
\renewcommand\@biblabel[1]{#1}            
\renewcommand\@makefntext[1]%
{\noindent\makebox[0pt][r]{\@thefnmark\,}#1}
\makeatother 
\renewcommand{\figurename}{\small{Fig.}~}
\sectionfont{\large}
\subsectionfont{\normalsize} 

\fancyfoot{}
\fancyfoot[RO]{\footnotesize{\sffamily{1--\pageref{LastPage} ~\textbar  \hspace{2pt}\thepage}}}
\fancyfoot[LE]{\footnotesize{\sffamily{\thepage~\textbar\hspace{3.45cm} 1--\pageref{LastPage}}}}
\fancyhead{}
\renewcommand{\headrulewidth}{1pt} 
\renewcommand{\footrulewidth}{1pt}
\setlength{\arrayrulewidth}{1pt}
\setlength{\columnsep}{6.5mm}
\setlength\bibsep{1pt}

\twocolumn[
  \begin{@twocolumnfalse}
\noindent\LARGE{\textbf{Elasto-capillarity at the nanoscale: on the coupling between elasticity and surface energy in soft solids}}\

\vspace{0.6cm}

\noindent\large{\textbf{Joost H. Weijs$^{\ast}$\textit{$^{a}$}, Bruno Andreotti\textit{$^{b\ddag}$}, and  Jacco H. Snoeijer\textit{$^{a}$}}}\vspace{0.5cm}

%

\vspace{0.6cm}

\noindent \normalsize{The capillary forces exerted by liquid drops and bubbles on a soft solid are directly measured using molecular dynamics simulations. The force on the solid by the liquid near the contact line is not oriented along the liquid vapor interface nor perpendicular to the solid surface, as usually assumed, but points towards the liquid.
 It is shown that the elastic deformations induced by this force can only be explained if, contrary to an incompressible liquid, the surface stress is different from the surface energy. Using thermodynamic variations we show that the the surface stress and the surface energy can both be determined accurately by measuring the deformation of a slender body  plunged in a liquid. The results obtained from molecular dynamics fully confirm those recently obtained experimentally [Marchand \emph{et al.} Phys. Rev. Lett. {\bf 108}, 094301 (2012)] for an elastomeric wire. 
}
\vspace{0.5cm}
 \end{@twocolumnfalse}
  ]

\section{Introduction}


\footnotetext{\textit{$^{a}$~Physics of Fluids Group, Faculty of Science and Technology and Mesa+ Institute,
University of Twente, P.O. Box 217, 7500 AE Enschede, The Netherlands.}}
\footnotetext{\textit{$^{b}$~Physique et M\'ecanique des Milieux H\'et\'erog\`enes, UMR
7636 ESPCI -CNRS, Univ. Paris-Diderot, 10 rue Vauquelin, 75005, Paris,
France}}

As largely demonstrated in the last two decades, elasticity plays an important role in surface physics. Phenomena such as surface reconstruction \cite{BGIE97,FF97}, surface segregation \cite{WK77}, surface adsorption \cite{I04}, elastic instabilities \cite{MS04}, self assembly \cite{AVMJ88,MSM05}, and nanostructuration \cite{R02} of crystalline solids are directly induced by surface stresses. In parallel, and almost without any connection, the elastic deformations of sheets and rods\cite{BicoNATURE,PRDBRB07,BoudPRE,PyEPJST,HonsAPL,RomanJPCM,ChiodiEPL,HurePRL} as well as gels and elastomers induced by capillary forces have been evidenced and investigated \cite{L61,Rusanov75,Yuk86,Shanahan87,CGS96,White03,PericetCamara08,PBBB08,MPFPP10,SARLBB10,LM11,JXWD11,DMAS11,SBCWWD13,WBZ09,Style12,Limat12}. 
It has remained unclear to what extent it is important to distinguish surface tension from surface stress for these elasto-capillary phenomena.

The definition and properties of surface stresses can be derived from thermodynamics, atomistic models and mechanics. These approaches are complementary and should in principle be consistent with one another. The simplest situation is an interface between a condensed phase and its vapor. Let us consider an extensive quantity, the density of which varies across the interface over the scale of a few molecular sizes. At a macroscopic scale, the density can be seen as homogeneous on both sides of the interface. However, the extensive quantity then presents an interfacial excess. For instance, the free energy presents an interfacial excess called the ``surface energy", denoted $\gamma$. Dividing now the total volume of the condensed phase and its vapor into two subsystems with a dividing plane \emph{normal} to the interface, the mechanical force between the two subsystems also presents an excess quantity called the ``surface stress", which throughout the paper we refer to as $\Upsilon$. This is a force per unit length acting parallel to the interface, originating from molecular interactions. 

In the very particular situation where the condensed phase is an incompressible liquid, it can be shown from the virtual work principle that the surface stress and the surface energy are strictly equal, i.e. $\Upsilon=\gamma$. The surface stress and surface energy are then unified into a single name ``surface tension", and it is common to address capillary problems using either the thermodynamic, or the mechanical route \cite{bookDeGennes,MWSA11}. In a solid, by contrast, the surface energy a priori depends on the strain in the bulk and yields an additional elastic contribution to the surface stress \cite{Shuttleworth50, MS04}. More precisely, the difference between the surface stress and the surface energy is the derivative of the surface energy with respect to the strain. This result is known as the Shuttleworth equation \cite{Shuttleworth50, MS04}, 

\begin{equation}\label{eq:shuttleworth}
\Upsilon_{ij} = \gamma_{ij} + \frac{\partial \gamma_{ij}}{\partial \epsilon},
\end{equation}
where $\epsilon$ is the bulk strain parallel to the interface. The subscripts refer to the phases $i$ and $j$ on both sides of the interface. Indeed, the Shuttleworth equation also applies when the interface separates two condensed phases composed of different molecules. This is highly relevant for wetting phenomena, for which one naturally deals with liquid-solid interfaces. Such interfaces present an excess free energy $\gamma_{SL}$ that, according to (\ref{eq:shuttleworth}), is different from the surface stress $\Upsilon_{SL}$. Once again, the surface stress is the excess force parallel to the interface and can be measured at the edge of any control volume that includes the interface. Importantly, the force $\Upsilon_{SL}$ is exerted on a subsystem composed two types of molecules, solid and liquid. From a molecular perspective, $\Upsilon_{SL}$ is the resultant of all types of molecular forces: solid-liquid, solid-solid and liquid-liquid interactions. 
\begin{figure*}[h!t]
\centering
  \includegraphics{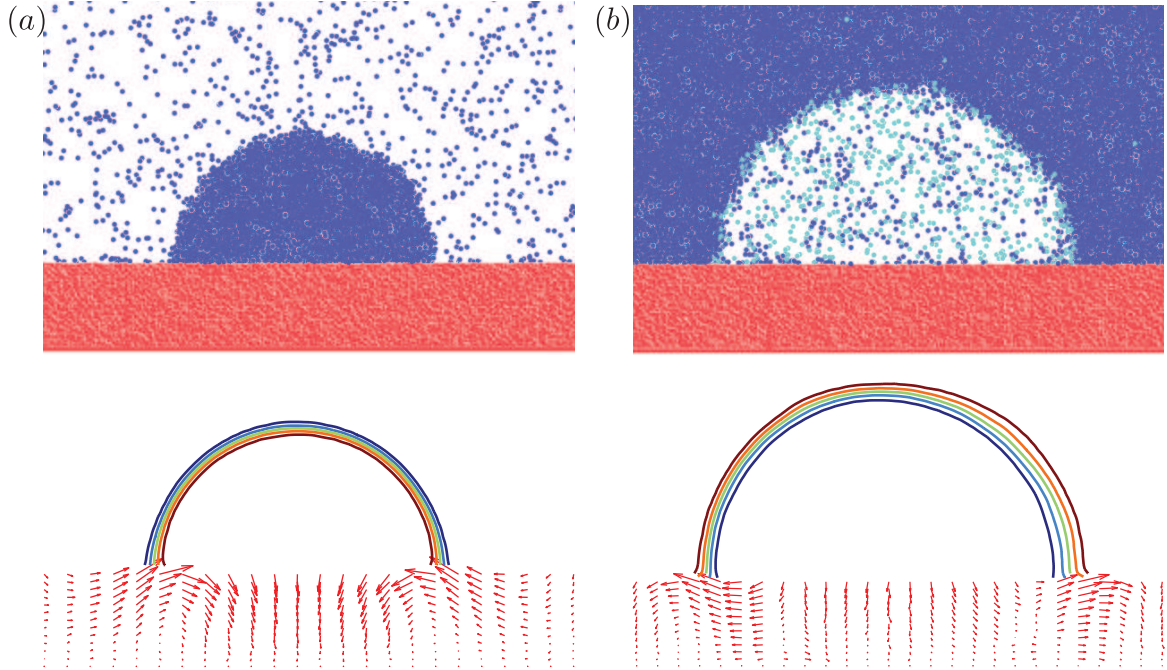}
  \caption{Molecular dynamics simulations of a drop (a) and a bubble (b) on a deformable substrate, both with contact angle $\theta=90^\circ$, hence $\gamma_{SL}=\gamma_{SV}$. (a) Drop on a soft substrate. {\em Top}: Snapshot, blue particles are liquid atoms, red particles are solid atoms. {\em Bottom}: Local displacement (red arrows) of the solid due to the presence of the liquid drop (shown by liquid isodensity contours). (b) Bubble on a soft substrate. {\em Top}: Snapshot, red and blue particles are the same as in (a), cyan particles are gas atoms. {\em Bottom}: Local displacement (red arrows) of the solid due to the presence of the bubble (shown by liquid isodensity contours).
 Note that the tangential displacement near the contact line is different between both situations: in the drop case the solid is pulled inwards whereas in the bubble case the solid is pulled outwards.}
  \label{Fig1}
  \end{figure*}

The aim of this paper is to explore the difference between surface energy and surface stress in the case of soft solids that are partially wetted by a liquid. A paradigmatic example of this effect consists of a drop (or bubble) on a deformable solid \cite{L61,Rusanov75,White03,PericetCamara08,PBBB08,WBZ09,DMAS11,JXWD11,Style12,Limat12,SBCWWD13}, as shown in Fig.~\ref{Fig1}. This problem has recently been explored experimentally and theoretically, with contradicting interpretations emerging from microscopic and macroscopic descriptions \cite{PBBB08,LM11,JXWD11,Style12,SBCWWD13,DMAS11,MDSA12b,Limat12}. Indeed, we expect a particularly strong manifestation of the difference between surface energy and surface stress near a three-phase contact line. On the one hand, the liquid-vapor interface is characterized by a surface tension, i.e. $\gamma_{LV}=\Upsilon_{LV}$, and Young's law for the equilibrium contact angle 
\begin{equation}
\gamma_{LV} \cos\theta = \gamma_{SV} - \gamma_{SL}
\label{eq:young}
\end{equation}
only involves the surface energies. On the other hand, the mechanical equilibrium of the solid involves the surface stresses $\Upsilon_{SV}\neq \gamma_{SV}$ and $\Upsilon_{SL}\neq \gamma_{SL}$. 

In this work, we reveal the connection between elasto-capillary interactions at the nanoscale and the thermodynamic concepts of surface energy and surface stress. In Section \ref{sec:drop}, we perform Molecular Dynamics simulations of drops and bubbles on soft substrates, for which we can accurately determine the elastic deformations and the liquid-on-solid forces. We find that the deformation below a drop is markedly different from the deformation below a bubble (Fig.~\ref{Fig1}): we measure a tangential force oriented towards the liquid side, even for contact angles of $\theta=90^\circ$ for which $\gamma_{SV}=\gamma_{SL}$. Can one explain this tangential force by invoking surface stresses? In Section \ref{sec:plate} we therefore develop a purely thermodynamic view, in the case of a plate partially immersed in a liquid, and compare this directly to Molecular Dynamics simulations. Finally, we conclude in Section~4 by relating the tangential forces to the difference between surface energies and stresses.  

\section{Drops and bubbles}\label{sec:drop}
\subsection{Molecular Dynamics}\label{sec:MDdetails}
The aim of this paper is to investigate the basic mechanisms controlling elasto-capillary interactions in model situations. We have therefore performed Molecular Dynamics simulations using simple interactions for both the liquid and the soft solid, in a quasi-2D geometry. The simulations have been performed using the {\sc Gromacs} software package\cite{gromacs}. The liquid consists of a Lennard-Jones fluid whose pair interaction potential is given by:
\begin{equation}
\phi_{ij}(r) = 4e_{ij}\left[\left(\frac{d_{ij}}{r}\right)^{12}-\left(\frac{d_{ij}}{r}\right)^{6}\right]\;.
\end{equation}
Here, $e_{ij}$ and $d_{ij}$ are the interaction strength and range between particle types $i$ and $j$, respectively. The potential is cut off at $5d_{LL}$, where $d_{LL}$ is the liquid atom size. The simulations are performed in the $NVT$-ensemble (constant number of particles $N$, constant volume $V$, and constant temperature $T$ using the thermostat described in \cite{Bussi07}). 
The solid consists of atoms placed on a cubic lattice of 25 layers, with harmonic springs connecting each atom to its neighbour and next-nearest neighbour. The lattice spacing $a=0.8 d_{LL}$, while all spring constants are taken equal with $k=38.5\cdot e_{LL}/d_{LL}^2$. There are no Lennard-Jones solid-solid interactions, hence the solid atoms only interact with each other through the harmonic springs. The solid (S) interacts with the liquid (L) through Lennard-Jones interactions. By varying the solid-liquid interaction, we explore drops of different equilibrium contact angles \cite{WMALS11}. In the simulations for a gas bubble, we added gas atoms (G) that also interact according to a Lennard-Jones potential.
The addition of gas atoms in the bubble case is required to prevent the bubble from collapsing immediately, as would be the case for a vapour bubble.
The Lennard-Jones interaction parameters are given in Table~\ref{tbl:interactions}. In the following paragraph we define the relevant dimensionless numbers.

From a macroscopic view, the elasto-capillary deformations arise from a balance between surface tension $\gamma$ and the elastic modulus $E$ (for simplicity of notation we use $\gamma$ for the liquid-vapor surface tension $\gamma_{LV}$). The ratio of these parameters $\gamma/E$ gives the elastocapillary length, which sets the scale of the elastic deformations. Our quasi-two-dimensional simulations have plane-strain conditions, in which case the relevant elastic modulus reads $E=\tilde{E}/(1-\nu^2)$, where $\tilde{E}$ is the Young's modulus and $\nu$ the Poisson ratio. In terms of lattice parameters in our simulations, we find $E=15k/(8a)$. 
To quantify the relative softness of the substrate, one can compare the elastocapillary length to the (liquid) atomic size $d_{LL}$, which gives the dimensionless quantity $\gamma/(Ed_{LL})$. Whenever this quantity is small, the deformations are weak and one should recover the contact angles according to Young's law \cite{MDSA12b}. 
Here we measure the liquid-vapour surface tension $\gamma$ in a separate system using a Kirkwood-Buff integral over the stress-anisotropy near the interfaces  $\gamma=\int (p_N-p_T(z))dz$ where, $p_N$ is the (constant) thermodynamic pressure in the system, and $p_T$ the tangential (relative to the interface) component of the stress-tensor which deviates from $p_N$ near the liquid-vapour interface \cite{KB1949,Nijmeijer,WMALS11}. We find, using the method outlined in \cite{Nijmerijer} to determine the local pressure that $\gamma=0.78 \; e_{LL}/d_{LL}^2$.
In this work, therefore, the parameter $\gamma/(Ed_{LL})=8.6\cdot 10^{-3}$ is indeed small, meaning that all elastic displacements are much smaller than atomic size (small elastic strains), and allows the use of linear elasticity theory.
The typical length scale of thermal fluctuations in the solid, $\sqrt{k_BT/k}$, compared to the elastocapillary deformation described before provides another dimensionless quantity $\frac{\gamma \sqrt{k} }{E\sqrt{k_BT}}=0.18$. The smallness of this parameter shows that the thermal fluctuations in the system are much larger than the deformation due to capillary forces. Still, as we will show, the displacement field can be measured very accurately in the simulations after averaging over time. Finally, we note that in SI-units, the chosen material properties correspond to real materials at $T=300$K, $E=11$~GPa, and $\gamma=3.1\cdot 10^{-2}$~J/m$^2$.
Surface tension coefficients of simple liquids typically lie between $0.02$ J/m$^2$ (ethanol) and $0.07$ J/m$^2$ (water). Young's modulus of crystalline solids is typically around  $E=100$~GPa while it can be much lower for elastomers (between $10^-2$ and $10^-1$ GPa for rubber) and gels (down to $1$~kPa) whose elasticity is entropic.
\begin{table}[t!]
\small
  \caption{\ Lennard-Jones Interaction parameters for the MD-simulations for Liquid, Solid and Gas atoms. With these valus, the liquid-vapour surface tension is $\gamma=3.1\cdot 10^{-2}$~J/m$^2$.}
  \label{tbl:interactions}
  \begin{tabular*}{0.4\textwidth}{@{\extracolsep{\fill}}llll}
    \hline
    Interaction pair $i,j$ & $\frac{e_{ij}}{k_BT}$ ($T=300$K)&  $d$/nm \\
    \hline
    LL & 1.2 &   0.34  \\
    SL & varied &    0.34\\
    SS & 0 &     0\\
    GG & 0.4 &    0.5\\
    SG & 0.004   &0.34 \\
    LG & 0.7 &    0.42\\

    \hline
  \end{tabular*}
\end{table}


The small strains in the solid (smaller than thermal fluctuations) are measured by calculating the time-averaged displacements (relative to the center of mass of the droplet or bubble) of the solid atoms compared to a base state, obtained from a simulation of the same solid in vacuum. As we are interested in the influence of the liquid on the solid, this procedure allows us to exclude effects associated with the presence of a solid-vacuum interface. The contact angle of the droplet and of the bubble are measured by determining the time-averaged density field of the liquid and the position of the Gibbs interface. We refer the reader to our previous work~\cite{WMALS11} for technical details. The circular fit to this liquid-vapor/gas interface is extrapolated to the solid, which provides the contact angle. 

\subsection{Elasto-capillary deformations}
To illustrate that surface energy is not sufficient to characterize elasto-capillary deformations, we first consider a case where $\gamma_{SV}= \gamma_{SL}$, such that the contact angle of the liquid is close to $90^\circ$. The contact angle can be adjusted by tuning only the Lennard-Jones interaction $e_{SL}$, while keeping the liquid parameters ($e_{LL}$, $\sigma_{LL}$, $\sigma_{SL}$) fixed. 
This way, the liquid properties are unchanged except for the interaction with the solid.
Note that, in general, the surface stresses will be different from the surface energies, and thus $\Upsilon_{SV}\neq \Upsilon_{SL}$. 

Figure \ref{Fig1} shows the elastic deformation in the solid below a liquid drop (panel a), and below a bubble filled with gas (panel b). Since $\theta=90^\circ$, the shape of the liquid-vapor interface, characterized by the iso-density profiles, is very similar in both cases. By contrast, the elastic deformations are markedly different, as can be seen from the vector field (red arrows): while below the drop one observes a tangential displacement towards the center of the drop, the displacements below the bubble are oriented outwards. This surprising outcome has important consequences. The drop and the bubble are perfectly symmetric from the point of view of the surface energies, since $\gamma_{SV}=\gamma_{SL}$ in this case. Yet, this symmetry is not reflected in the surface displacements: the deformations are not invariant under an inversion of the phases. Instead, the solid is always pulled towards the liquid side of the contact line (not only for the case $\theta=90^\circ $). Therefore, an elasto-capillary description based on constant surface energies (i.e. on surface tensions) is not sufficient to describe the elastic deformations below a drop or below a bubble. 


\subsection{Capillary traction and contact line force}

\begin{figure*}[h!t]
\centering
  \includegraphics{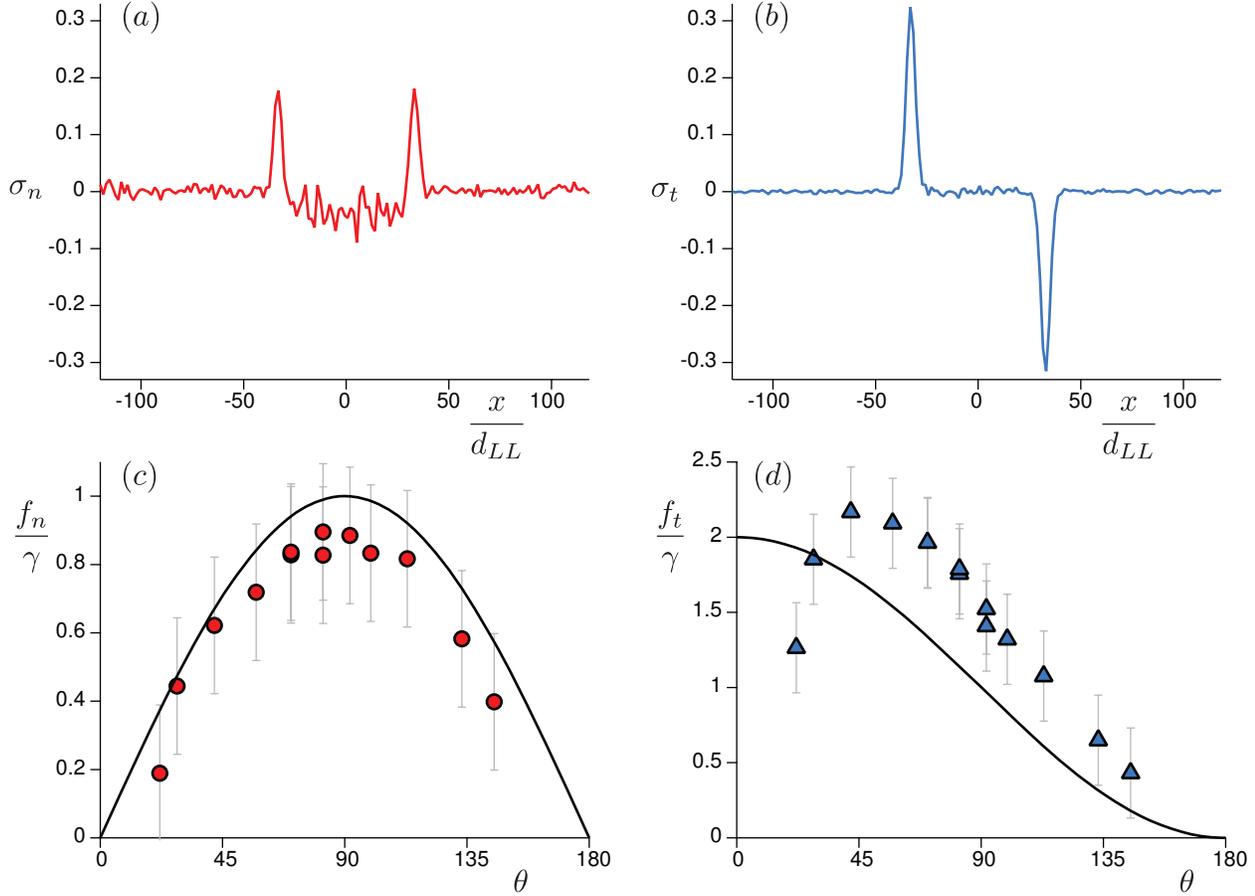}
  \caption{Capillary traction: the liquid on solid forces per unit area of the substrate measured in molecular dynamics of a droplet on a substrate ($\theta = 82^\circ$).   (a) Normal component of the force per area by the liquid on the solid, $\sigma_n$. The two peaks correspond to the contact lines, where the solid gets pulled up. The region between the peaks corresponds to the liquid-solid interface, where the solid gets pushed down due to the Laplace pressure.   (b) Tangential component of the force per area by the liquid on the solid, $\sigma_t$. There is only a force near the contact lines, and the force is directed towards the interior of the droplet.   (c) Normal component of the total force by the liquid on the solid due to the contact line at varying $\theta$. The solid line corresponds to $\sin\theta$.    (d) Tangential component of the total force by the liquid on the solid due to the contact line at varying $\theta$. The solid line corresponds to $1+\cos\theta$. The low values for small contact angles $\theta<40^\circ$ are likely due to finite size effects in the simulations.}
\label{Fig3}
\end{figure*}

Before turning to a fully thermodynamic description of elasto-capillary deformations in Sec.~\ref{sec:plate}, we first quantify the capillary liquid on solid forces at the nanoscale. 
Note that for a complete description of the bulk deformations one requires both liquid on solid forces {\em and} any solid on solid forces that are present in the surface layer. However, since the asymmetry is caused by the liquid on solid forces we first quantify the liquid on solid forces only.
The common feature of the deformation below the drop and the bubble is that, below the contact line, the tangential deformations are oriented towards the side of the liquid phase (Fig.~\ref{Fig1}). This is consistent with the predictions of the Density Functional Theory in the sharp interface approximation \cite{MWSA11,DMAS11}, which is based on a microscopic description of the interactions. The mechanism for this asymmetry is that the long-ranged attraction by the liquid molecules creates a resultant force on the solid that is biased towards the liquid: the solid is more strongly attracted by the phase of highest density. This resultant force ultimately determines the elastic deformations, and is responsible for breaking the symmetry between drops and bubbles in Fig. \ref{Fig1}. 

In Molecular Dynamics, we can of course directly quantify this effect by measuring the time-averaged forces that all liquid molecules exert on the solid molecules. Since we wish to reveal the capillary traction (force per area) that the liquid induces on the solid, we integrate over the vertical depth of the solid. In practice, the direct liquid-solid interaction only acts in the superficial layers of the solid, due to the short range of Lennard-Jones (van der Waals) interactions ($\sim r^{-6}$). The resulting capillary traction below a drop is shown in Fig.~\ref{Fig3}. First, the curve in Fig.~\ref{Fig3}a shows the normal traction, $\sigma_n$, of the liquid on the solid in the case $\theta=82^\circ$. As expected, we observe a large traction that is localized near the two contact lines: this corresponds to the ``pulling'' action of the contact line. The width of the peak is a few molecular sizes and reflects the width of the liquid-vapor interface. At the center of the drop one observes a slightly negative traction, corresponding to the Laplace pressure in the bulk of the drop. As the drop is in equilibrium, this Laplace pressure perfectly balances the upward stress at the contact line: the total liquid-on-solid normal force is zero. 
Using macroscopic thermodynamics we can estimate the normal force exerted by the liquid on the solid in the vicinity of the contact line. For drop sizes that are much larger than the width of the peak in $\sigma_n$ (which coincides with the thickness of the liquid-vapour interface), we can separate the capillarity forces into a contribution per unit area, $P = \gamma/R$, and a perfectly localized force per unit contact line $f_n$. Using that the width of the drop $2R\sin\theta$ and that the total force vanishes, one predicts
that the strength of the force on the solid near the contact line should be $f_n= \gamma \sin \theta$ (per unit contact line).
We can test this macroscopic prediction in our simulations. First, we determine the liquid-vapor surface tension $\gamma$ from an independent calibration, as described in Sec.~\ref{sec:MDdetails}. Then, we integrate the normal stress over the peak located around the contact line, yielding the total normal force per unit contact line $f_n$. The result for different contact angles $\theta$ is shown in Fig.~\ref{Fig3}c. The results of MD simulations are consistent with a contact line force in the normal direction $f_n = \gamma \sin \theta$, shown in solid line.

Similarly, we can determine the tangential capillary traction on the solid, denoted $\sigma_t$. The result is shown in Fig.~\ref{Fig3}b. We find a positive traction at the contact line located on the left (i.e. pointing towards the right), and a negative traction at the contact line located on the right (i.e. pointing towards the left). Indeed, we identify an effective liquid-on-solid force that is oriented towards the interior of the drop, i.e. into the liquid phase. This tangential force is the reason why the elastic displacements point towards the interior of the drop. Again, we quantify the total force exerted on the solid near the contact line from the integral of the peaks, $f_t$. The resulting tangential force per unit contact line is shown in Fig.~\ref{Fig3}d, as a function of the contact angle. We observe a nonzero inward tangential force for all angles. This can be understood from the left-right symmetry breaking below the contact line: the solid atoms are attracted much more strongly by the high-density liquid phase than by the low-density vapor phase. As the strength of the solid-liquid interaction is directly quantified by the work of adhesion, $\gamma+\gamma_{SV}-\gamma_{SL}$, one expects below a liquid/vapour/solid triple line \cite{MWSA11}$^,$\footnote{In case the vapor is replaced by a second, immiscible liquid, there will be second ``work of adhesion'' contribution. This yields another tangential force on the solid that is biased toward the second liquid, and weakens the asymmetry.}:
\begin{equation}
f_t = \gamma + \gamma_{SV} - \gamma_{SL} = \gamma\;(1+\cos \theta)\;.
\end{equation}
This equation is shown as the solid line in Fig.~\ref{Fig3}d: it indeed captures the features of the tangential force, which is always positive, i.e. oriented towards the liquid side, and nicely describes the magnitude and trend with the contact angle. This also explains the difference between the deformations below a drop and a bubble. 

In conclusion, our simulations clearly demonstrate the existence of a tangential capillary force exerted on the region of the solid below  the contact line. This force has a strong influence on the elastic deformation below a drop or bubble. This effect is usually ignored in the literature on elasto-capillarity \cite{L61,Rusanov75,White03,JXWD11,Style12,SBCWWD13}, likely due to the fact that in the case of an incompressible liquid (or solid) this tangential force is exactly balanced by solid on solid forces and therefore is not transmitted to the bulk. However, in general for $\nu\ne 1/2$, this is not the case and the tangential force (which is always pointed towards the liquid phase, see Fig.~\ref{Fig3}d) needs to be taken into account when considering the elastic deformation.

\section{Partially immersed solid}\label{sec:plate}
From the preceding section is it clear that the elastic deformation below a contact line results from the detailed interactions (capillary and elastic) at the nanoscale. We will now address the problem in a macroscopic framework, where we relate the elastic displacements to purely thermodynamic concepts. In particular, the goal is to express the microscopic interactions discussed in Sec. 2 directly in terms of the excess quantities $\gamma_{ij}$ and $\Upsilon_{ij}$.
For this, we consider an long elastic plate that is partially immersed in a liquid -- see Fig.~\ref{thermosketch}. This geometry was studied before experimentally using a thin elastomeric wire \cite{MDSA12}. The experiment revealed that the elastic strain in the ``wet'' part of the solid was very different from the strain in the ``dry'' part. Here we analyze this geometry using the thermodynamic concepts of surface stress and then compare it directly to Molecular Dynamics. 

The central result of this section is that the vertical strain above the contact line, $\tilde{\varepsilon}_+$, couples to the surface energies, while the strain below the contact line, $\tilde{\varepsilon}_-$, is determined by the surface stresses:

\begin{eqnarray}\label{eq:strainup}
\frac{WE}{2}\, \tilde{\varepsilon}_+ &=& \gamma_{SV}-\gamma_{SL}\\
\frac{WE}{2}\, \tilde{\varepsilon}_- &=& \Upsilon_{SV}-\Upsilon_{SL} \label{epsmin} \;.\label{eq:straindown}
\end{eqnarray}
Here $W$ is the width of a two-dimensional elastic plate, $E$ is the elastic modulus, while the reference state for the strain is the completely dry solid. By determining the strain inside the plate, one thus directly measures the difference between surface energies and surface stresses. This method will be applied in our Molecular Dynamics simulations. 

\subsection{Thermodynamics: strain above and below meniscus}
\begin{figure}[ht]
\centering
\includegraphics{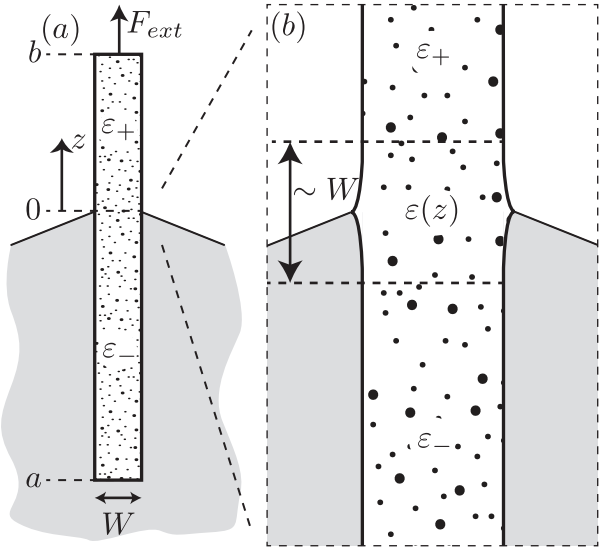}
  \caption{Partially immersed plate of width $W$, held at equilibrium by an external force $F_{ext}$. (a) The plate is  partially wetted, with the contact line located at $z=0$. There is a homogeneous strain above and below the contact line: $\varepsilon_+$ and $\varepsilon_-$, respectively. (b) Zoom around the contact line. Near the contact line exists a transition region of the strain from $\varepsilon_+$ to $\varepsilon_-$. The transition occurs over a length $\sim W$.}
  \label{thermosketch}
\end{figure}

To derive (\ref{eq:strainup}) and (\ref{eq:straindown}), we consider the free energy of an elastic plate that is partially submerged in a liquid bath with the contact line at $z=0$ (Fig.~\ref{thermosketch}). We will assume that the elasto-capillary length $\gamma/E$ is much smaller than the plate width $W$, which in turn is much smaller than the total plate length $L$:

\begin{equation}
\frac{\gamma}{E} \ll W \ll L~.
\end{equation}
The first assumption allows one to consider the solid-liquid and solid-vapor interfaces to be flat with respect to other scales in the problem, as can be seen from the zoom near the contact line in Fig.~\ref{thermosketch}b. The second assumption implies that, apart from the region directly below the contact line of width $W$, the strain is homogeneous and characterized by a constant value for $\epsilon= du_z/dz$, where $u_z$ is the vertical displacement field. We denote the strains above and below the contact line as $\epsilon_+$ and $\epsilon_-$ respectively; the contribution to the energy of the region directly below the contact line is sub-dominant by a factor $W/L$. Outside the contact line region the strains can be considered small, since $\epsilon_{\pm} \sim \gamma/EW \ll 1$. Under these assumptions, we arrive at the following free energy functional of the system (per unit length):

\begin{eqnarray}\label{eq:free}
\mathcal{F} &=& 
2 \int_a^{z_{cl}} dz \, \left[ \gamma \left(1+h'^2 \right)^{1/2} +\gamma_{SL} \right] 
+2 \int_{z_{cl}}^b dz \, \gamma_{SV} \nonumber \\
&&
+ W \int_{a_0}^{z_{cl}}dz\, \frac{1}{2} E \epsilon_-^2 + W \int_{z_{cl}}^{b_0}dz\, \frac{1}{2} E \epsilon_+^2.
\end{eqnarray}
Here $a$ and $b$ denoted the bottom and top positions of the plate, but note that the elastic energy should be taken over a domain of fixed length $L_0=b_0-a_0$, with reference positions $a_0$ and $b_0$ \cite{LandauLifshitz}. We allow for variations of the shape of the liquid-vapor interface $h(z)$, the position of the contact line $z_{cl}$, the top and bottom positions of the plate $b$ and $a$, and the elastic strains $\epsilon_\pm$. The equilibrium conditions follow from 
\begin{equation}
\delta \mathcal{F}= F_{ext}\delta b,
\end{equation}
which equates the change in energy to the work done by the external force. Note that the positions $b$ and $a$ are linked by the constraint
\begin{equation}\label{eq:constraint}
b-a = L_0 + \int_{a_0}^{b_0} dz\; \epsilon.
\end{equation}

First, we derive the equilibrium conditions for the liquid, by considering variations of the liquid-vapor interface $\delta h(z)$, with $\delta b = \delta \epsilon_+=\delta \epsilon_- =0$. From geometry near the contact line, this implies a variation of the contact line position according to $\delta h(z_{cl}) =- h'(z_{cl})\delta z_{cl}$. One thus obtains 

\begin{eqnarray}
&&\frac{1}{2} \delta\mathcal{F} = 0=\nonumber \\ 
&&\delta z_{cl} \left[ \gamma \left(1+h'^2 \right)^{1/2} + \gamma_{SL} - \gamma_{SV} +\frac{1}{2}EW \left(\epsilon_-^2 - \epsilon_+^2 \right) \right]_{z_{cl}} \nonumber \\
&& +\delta h \left[ \frac{\gamma h'}{\left(1+h'^2 \right)^{1/2}}\right]_{z_{cl}} 
-  \int dz\; \frac{\gamma h''}{\left(1+h'^2 \right)^{3/2}}\, \delta h.
\end{eqnarray}
The integral expresses the Laplace pressure condition for the liquid-vapor interface. The terms $\sim \epsilon_\pm^2$ arise from the fact that a variation of $z_{cl}$ does not affect the contact line zone, but just gives an exchange of the elastic energies of the dry and wet parts respectively (similar to the exchange of surface energies $\gamma_{SL}-\gamma_{SV}$). Collecting the terms from the boundary condition, using $\delta h(z_{cl}) =- h'(z_{cl})\delta z_{cl}$, one finds the condition for the equilibrium contact angle

\begin{eqnarray}
\cos \theta &=& \frac{\gamma_{SV} -\gamma_{SL}}{\gamma} + \frac{1}{2}\frac{EW}{\gamma} \left(\epsilon_+^2 - \epsilon_-^2\right) \nonumber \\
&=& \frac{\gamma_{SV} -\gamma_{SL}}{\gamma} + \mathcal{O} \left( \frac{\gamma}{EW}\right).
\end{eqnarray}
where we replaced $\cos \theta = 1/(1+h'^2)^{1/2}$. This shows that for $\gamma/EW \ll 1$ one recovers Young's law for the liquid contact angle with respect to the undeformed solid. 

Next, we explore the elastic degrees of freedom of the plate. For convenience, we now choose the contact line position as the reference altitude: $z_{cl}=0$. Using Eq.~(\ref{eq:free}), one writes the free energy $\mathcal{F}_p$ of the plate and its interfaces with the liquid and the vapor:
\begin{eqnarray}
\mathcal{F}_p = 2b\gamma_{SV}(\epsilon_+)\ - 2a \gamma_{SL}(\epsilon_-) 
+ \frac{1}{2}WE \left(  b_0 \epsilon_+^2  - a_0 \epsilon_-^2\right).
\end{eqnarray}
Here we made explicit the strain dependence of surface energies, which is necessary for solid interfaces. Due to the relations
\begin{equation}
b =b_0\left(1+ \epsilon_+ \right), \quad a=a_0\left(1+\epsilon_-\right), \quad a_0=b_0 - L_0,
\end{equation}
there are only three independent variables. We choose here to parametrize the problem using $b_0$, $\epsilon_+$ and $\epsilon_-$. From this we can write the total variation:
\begin{eqnarray}\label{eq:variation}
 \frac{1}{2}\delta\mathcal{F}_p &=&
 \delta b_0 \left[ \gamma_{SV} - \gamma_{SL} + \frac{1}{2} EW \left(\epsilon_+^2 - \epsilon_-^2 \right) \right] 
\nonumber \\
&&+ b_0 \delta \epsilon_+ \left[ \gamma_{SV} +\frac{\partial \gamma_{SV}}{\partial \epsilon} + \frac{EW}{2} \epsilon_+ \right]
\nonumber \\
&& - a_0 \delta \epsilon_- \left[ \gamma_{SL} +\frac{\partial \gamma_{SL}}{\partial \epsilon} + \frac{EW}{2} \epsilon_- \right],
\end{eqnarray}
where, after the variation, we replaced $b= b_0$ and $a=a_0$ owing to the smallness of the strains. For the same reason, we also anticipate that the terms of order $EW \epsilon_\pm^2$ can be neglected in the following steps. 

The variation of the plate energy must be balanced with the work done by the external force 

\begin{equation}\label{eq:work}
\delta \mathcal{F}_p = F_{ext} \delta b = F_{ext} \delta b_0 + F_{ext} b_0 \delta \epsilon_+.
\end{equation}
Combining (\ref{eq:variation}) and (\ref{eq:work}), gives the three equilibrium conditions. The variation of $b_0$ gives the familiar expression for the external force needed to hold the plate, i.e. 

\begin{equation}
F_{ext} = 2\left(\gamma_{SV} - \gamma_{SL}\right).
\end{equation}
Using this expression, we find that the strains must follow
\begin{eqnarray}
EW\varepsilon_+ &=&-2\frac{\partial \gamma_{SV}}{\partial \varepsilon}-2\gamma_{SL} \label{eq:epsp}\\
EW\varepsilon_- &=& -2\frac{\partial \gamma_{SL}}{\partial \varepsilon}-2\gamma_{SL} \label{eq:epsm}\;.
\end{eqnarray}
The reference state for the strain, $\varepsilon_0$, will be the plate in contact with the vapour only. This reference can be derived in a similar way by minimizing the free energy of a plate in vapour (i.e. replacing $\gamma_{SL}$ by $\gamma_{SV}$ in either of the above equations):
\begin{equation}
EW\varepsilon_0=-2\frac{\partial\gamma_{SV}}{\partial\varepsilon}-2\gamma_{SV}\;.
\end{equation}
The final step is to subtract the reference state from \eqref{eq:epsp} and \eqref{eq:epsm}, and substitute the Shuttleworth equation $\Upsilon_{ij}=\frac{\partial \gamma_{ij}}{\partial\varepsilon}+\gamma_{ij}$:
\begin{eqnarray}
\tilde{\varepsilon}_+ \equiv \varepsilon_+-\varepsilon_0&=& \frac{2}{EW}\left(\gamma_{sv}-\gamma_{sl}\right) \nonumber\\
\tilde{\varepsilon}_- \equiv \varepsilon_--\varepsilon_0&=& \frac{2}{EW}\left(\Upsilon_{sv}-\Upsilon_{sl}\right) \nonumber\;.
\end{eqnarray}
Indeed, this is the result anticipated in (\ref{eq:strainup}) and (\ref{eq:straindown}). In conclusion, the immersion of a slender body allows one to determine accurately both surface energies and surface stresses \cite{MDSA12}. 
\begin{figure}[t!]
\centering
\includegraphics{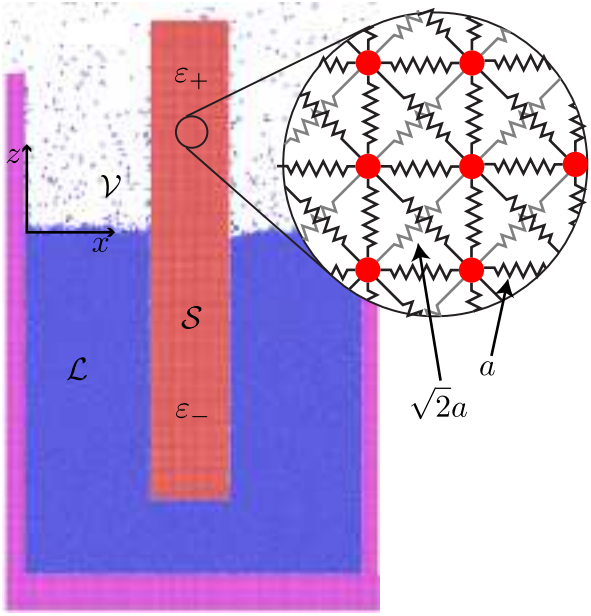}
  \caption{Snapshot from an MD-simulation of a plate ($W/a=26$ atoms wide) partially submerged in a liquid bath. The liquid-solid interaction energy $e_{SL}$ was chosen such that $\theta=90^\circ$. Comparing the time-averaged $z$-position of the solid atoms to the time-averaged position of the reference system (plate in vacuum), allows for the local displacement $u_z(z)$ to be measured.}
  \label{fig:platesnap}
\end{figure}

\subsection{Molecular Dynamics}
We now test this method  in a MD simulation of an immersed plate. Figure~\ref{fig:platesnap} shows a snapshot of the simulation, where the solid plate is partially immersed into a liquid reservoir. 
The material of the plate is the same as that used in the previous drop and bubble simulation, except that it is stiffer: $k_p=10k$, such that $E_p=110$~GPa whereas the liquid surface tension is unchanged: $\gamma=3.1\cdot 10^{-2}$~J/m$^2$.
The container that holds the liquid consists of the same material as the plate, except that the container is not allowed to deform by fixing the atoms to their initial positions. 
To avoid a curved liquid meniscus, and hence a difference in pressure on the wetted and dry regions of the plate, we consider again $\theta=90^\circ$. In this case $\gamma_{SV}= \gamma_{SL}$ and according to (\ref{eq:strainup}) we expect $\tilde{\varepsilon}_+=0$. Before each simulation we equilibrated the solid in a vacuum to have a well-defined reference state. 

\begin{figure}[t!]
\centering
\includegraphics{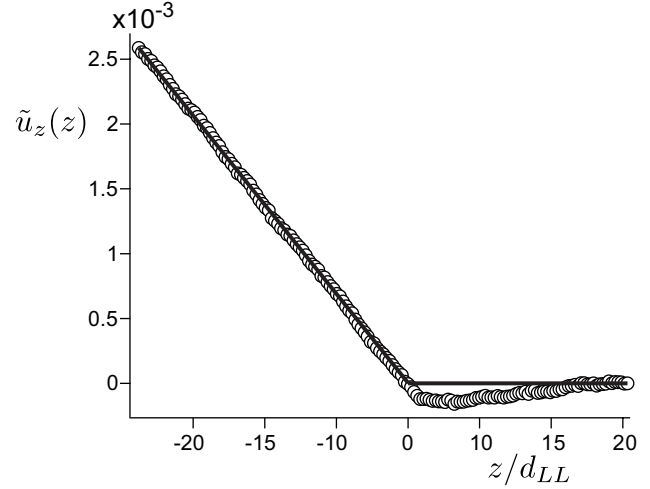}
  \caption{MD measurement of the relative displacement $\tilde u_z(z)$ for pa late width $W=8.8 d_{LL}$. The derivative of this slope gives the strain $\tilde \varepsilon(z)$. The contact line is located at $z=0$, where a clear jump in $\tilde\varepsilon(z)$ is observed (from $\tilde{\varepsilon}_-$ to $\tilde\varepsilon_+$, see also Figs.~\ref{thermosketch} and \ref{fig:platesnap}.) This signifies an imbalance in surface stresses ($\Upsilon_{SL}\ne\Upsilon_{SV}$) even though $\gamma_{SL}=\gamma_{SV}$ for $\theta=90^\circ$.}
  \label{Fig5}
\end{figure}

Figure~\ref{Fig5} shows an example of the vertical elastic displacement, $u_z(z)$, that is induced after immersion of the plate. The slope of this curve directly gives the strain $\tilde{\varepsilon}=du_z/dz$. Indeed, we observe very different strains above and below the contact line, which allows for a determination of $\tilde{\varepsilon}_+$ and $\tilde{\varepsilon}_-$. In this example the top part of the plate is hardly deformed, as expected from (\ref{eq:strainup}) for this situation where $\gamma_{SV}= \gamma_{SL}$. By contrast, the lower part of the plate displays a negative strain, $\tilde{\varepsilon}_- <0$, corresponding to a compression of the bottom part of the wire. Using (\ref{eq:straindown}), this reveals a difference in surface stresses, $\Upsilon_{SV}\neq \Upsilon_{SL}$, despite the equality of surface energies.

\begin{figure}[t!]
\centering
\includegraphics{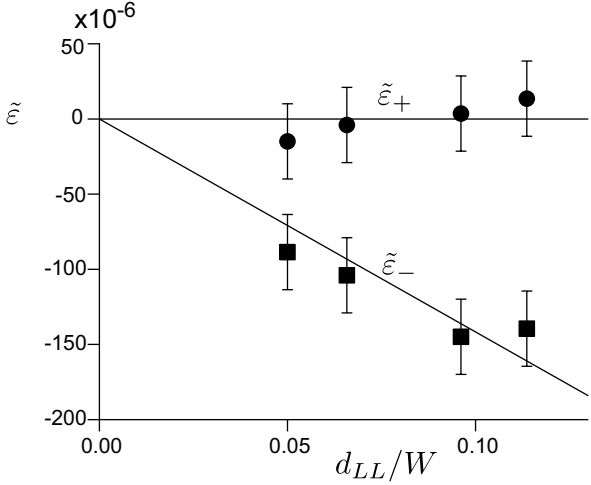}
  \caption{MD measurements of the vertical strains $\tilde\varepsilon_+$ (circles) and $\tilde\varepsilon_-$ (squares) as functions of the inverse plate width $d_{LL}/W$. Within error, there exists no strain above the contact line ($\varepsilon_+$). This is expected for $\theta=90^\circ$. Below the contact line, however, the solid is compressed due to an imbalance of the surfaces stresses at the contact line ($\Upsilon_{SL}\ne\Upsilon_{SV}$). The slope of this curve quantifies the magnitude of this imbalance, Eq.~\eqref{epsmin}.}
  \label{Fig6}
\end{figure}

To quantify the difference $\Upsilon_{SV}- \Upsilon_{SL}$, we repeated the simulations plates of various widths $W$. The resulting $\tilde{\varepsilon}_+$ and $\tilde{\varepsilon}_-$ are plotted as a function of $1/W$ (Fig.\ref{Fig6}). Above the contact line we indeed find a vanishing strain $\tilde{\varepsilon}_+$ for all plate thickness, within the error bars of the simulation. Below the contact line we observe a nearly linear dependence on $1/W$, as predicted by (\ref{eq:straindown}). This confirms that our system is large enough to apply thermodynamics and continuum elasticity. 
The slope of the curve is measured at $\frac{d\tilde\varepsilon_-}{d(1/W)}=-(1.4\pm 0.3)\cdot 10^{-3}\;d_{LL}$. This slope can be related to the surface stress via Eq.~\eqref{epsmin} and we find, for this specific solid and liquid pair:
\begin{equation}
\Upsilon_{SV}-\Upsilon_{SL} -(\gamma_{SV}-\gamma_{SL})= \Upsilon_{SV}-\Upsilon_{SL} = - (0.81 \pm 0.17) \;\gamma\;.
\end{equation}
This surface stress difference is clearly not a negligible effect: it is of the same order as the liquid-vapor surface tension. 

\section{Conclusions}

\begin{figure}[t!]
\centering
\includegraphics{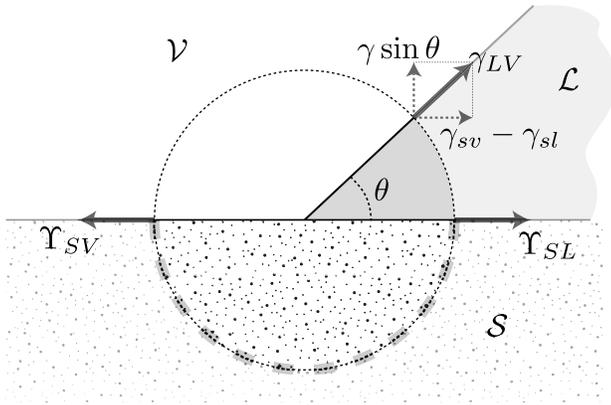}
  \caption{Stresses acting on the circular control volume around the contact line. Note that for solids generally $\Upsilon_{SX}\ne\gamma_{SX}$, hence there exists an imbalance of the interfacial stresses in both the normal and tangential directions. This imbalance is counteracted by elastic stresses in the solid, along the thick gray dashed line.}
  \label{fig:stressessketch}
\end{figure}

We have shown from thermodynamic considerations that the elastic deformation of a partially wetted solid crucially depends on the difference between surface stress and surface energy. This result is confirmed using Molecular Dynamics simulations, revealing how deformations emerge from interactions at the nanoscale. To complete the picture of elasto-capillary interactions, we finally give a purely mechanical interpretation of our findings. 

Describing the partially immersed wire of Fig.~\ref{thermosketch} using continuum elasticity, a discontinuity of strain implies a discontinuity of stress across the contact line, in the direction parallel to the solid interface. This means that the contact line region must exert a tangential force $f_t^{el}$ on the bulk elastic material \cite{MDSA12}. The magnitude of the tangential force experienced by bulk elasticity is proportional to $\epsilon_+ - \epsilon_-$, and therefore reads
\begin{equation}\label{eq:tang}
f_t^{el} = \left( \Upsilon_{SL} - \Upsilon_{SV} \right) - \left( \gamma_{SL} - \gamma_{SV}\right).
\end{equation}
This residual force accounts for \emph{all} interactions that are transmitted across the surface layers to the bulk elastic, including the solid-solid interactions. It is therefore important to distinguish $f_t^{el}$ from $f_t$ measured in Fig.~\ref{Fig3}: while the latter only included the liquid-on-solid forces, the surface stress captures the total excess surface force and includes all superficial interactions.

Figure~\ref{fig:stressessketch} shows how the residual force $f_t^{el}$ arises due to the imbalance of surface stresses in the vicinity of the contact line. When discussing the forces near the contact line, it is absolutely critical to explicitly specify the material system to which the forces are applied: a different choice of control volume will lead to different forces \cite{MWSA11,MDSA12}. Here we consider a macroscopic control volume that includes the three-phase contact line, as indicated by the dotted circle. As this includes the three interfaces, one can directly represents the surface stresses $\Upsilon_{SV}$, $\Upsilon_{SL}$ and $\Upsilon=\gamma$ as indicated by the solid arrows. Interestingly, the equilibrium contact angle does not involve the surface \emph{stresses} of the solid, but rather the surface \emph{energies} $\gamma_{SL}$ and $\gamma_{SV}$. This is not inconsistent with Fig.~\ref{fig:stressessketch}, because Young's law represents an equilibrium (a minimal free energy from the thermodynamic perspective and a balance of forces from the mechanical point of view) inside the liquid only, and thus requires a different control volume that does not include the solid~\cite{MWSA11}. As a consequence, the surface stresses in Fig.~\ref{fig:stressessketch} do not balance in the direction parallel to the solid, but yields a nonzero tangential force $f_t^{el} = \gamma \cos \theta + \Upsilon_{SL} - \Upsilon_{SV}$, in agreement with  (\ref{eq:tang}). Similarly, the surface stresses yield a resultant normal force 

\begin{equation}
f_n^{el} = \gamma \sin \theta.
\end{equation} 
To restore mechanical equilibrium inside the control volume, both $f_n^{el}$ and $f_t^{el}$ must be balanced by elastic stresses that are exerted along the along the grey dashed circular section in Fig.~\ref{fig:stressessketch}. The tangential component vanishes only when the surface energies and surface stresses are equal.

\begin{figure}[t!]
\centering
\includegraphics{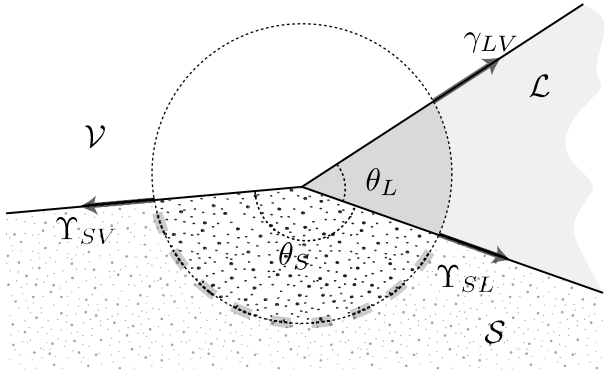}
  \caption{Stresses acting on the circular control volume around the contact line, in case of strong solid deformation ($\gamma/Ea \gg 1$). The liquid equilibrates at an angle $\theta_L$ that is a function of the solid angle $\theta_S$ and the surface energies $\gamma_{SL},\gamma_{SV}$. Since for solids generally $\Upsilon_{SX}\ne\gamma_{SX}$, the liquid equilibration does not coincide with a balance of the interfacial stresses. This imbalance is counteracted by elastic stresses in the solid, along the thick gray dashed line.}
  \label{fig:neumann}
\end{figure}

A similar observation can be made for very soft solids, for which the solid deforms into a ``cusp" shape with a solid angle $\theta_S < \pi$~\cite{MDSA12b,SBCWWD13,JXWD11}. The cusp develops when $\gamma/Ed_{LL} \gg 1$ \cite{MDSA12b}, and thus corresponds to cases much softer than in our Molecular Dynamics. Figure~\ref{fig:neumann} shows the surface stresses near the contact line on such a strongly deformed solid. For given value of $\theta_S$, the equilibration of the liquid angle $\theta_L$ involves only the surface energies $\gamma_{SL},\gamma_{SV}$ and not the stresses $\Upsilon_{SL},\Upsilon_{SV}$. In general, the liquid equilibrium will therefore not coincide with the balance of surface stresses on the circular control volume in Fig.~\ref{fig:neumann}\cite{MDSA12b}: residual elastic stress will arise whenever surface energies differ from surface stresses.

The key parameter for wetting of soft materials is thus the difference between surface energies and surface stresses, as in equation (\ref{eq:tang}). Experimentally, the strain discontinuity across the contact line for the partially immersed wire gives direct access to this difference. The recently suggested method to determine the surface stress from contact angles assumes a perfect balance of surface stresses \cite{SBCWWD13}, and therefore incorrectly assumes that elastic stress can be ignored for the balance that determines the contact angle. This method, therefore, applies only when there is no difference between surface energies and stresses. Theoretically, we can now put upper and lower bounds on the tangential force. Clearly, the simulations in Fig.~\ref{Fig1} show that the tangential force originates from the breaking of left/right symmetry near the contact line, biased towards the side of the high-density liquid. The maximum possible residual force $f_t^{el}$ should therefore be the liquid-on-solid force $\gamma + \gamma_{SV} - \gamma_{SL}$. This maximum arises whenever the solid-solid interactions in the surface layer do not counteract this effect and the full tangential force is transmitted to the bulk substrate. This was referred to as the ``vectorial force transmission model'' in previous work~\cite{MDSA12b}. Another extreme limit corresponds to the ``normal force transmission model", for which the surface layers completely screen out any tangential stress. This is the case, for example, when the substrate is another liquid (i.e. an oil drop floating on liquid). A liquid can of course not sustain any shear, which means that the symmetry-breaking of interactions is counteracted by self-interactions inside the liquid substrate. This once more agrees with (\ref{eq:tang}), since for liquid-liquid interfaces $\Upsilon_{ij}=\gamma_{ij}$. Hence, we conclude

\begin{equation}
0 \, \leq f_t^{el} \,\leq \, \gamma + \gamma_{SV} - \gamma_{SL},
\end{equation}
or equivalently

\begin{equation}
\gamma_{SL} -\gamma_{SV} \, \leq \, \Upsilon_{SL}-\Upsilon_{SV} \,\leq \  \gamma.
\end{equation}
Future work should be dedicated to determining how the difference between $\Upsilon$ and $\gamma$ exactly depend on the material dimensionless parameters: the solid Poission ratio the ratio of the elasto-capillary length to the atomic size and the ratio of the elasto-capillary length to the thermal length.
The results from our Molecular Dynamics are very close to the upper bound: the surface stress difference was found slightly smaller than the liquid-vapor surface tension $\gamma$. A similar conclusion can be drawn from the experimental results reported in \cite{MDSA12}, where an elastomeric wire that was partially immersed in a liquid. The symbol $\Gamma$ used in this previous work can now be identified with $\Upsilon_{SL}-\Upsilon_{SV}$, which was found identical to $\gamma$ within experimental uncertainty\cite{MDSA12}. To further explore the difference between surface energy and surface stress, it would be interesting to directly measure the tangential displacements inside soft substrates, e.g. using confocal microscopy \cite{SBCWWD13}, and compare the deformation below a drop and a bubble. 

{\bf Acknowledgments.} The authors gratefully acknowledge S. Das, L. Limat, D. Lohse and A. Marchand for many discussions. This work was sponsored by the NCF (Netherlands National Computing Facilities Foundation) for the use of supercomputer facilities and FOM, and STW (VIDI Grant No. 11304) all with financial support from the NWO (Netherlands Organization for Scientific Research).

\providecommand*{\mcitethebibliography}{\thebibliography}
\csname @ifundefined\endcsname{endmcitethebibliography}
{\let\endmcitethebibliography\endthebibliography}{}

\end{document}